\documentclass[aip,
 amsmath,amssymb,
 reprint,%
]{revtex4-1}

\usepackage{graphicx}% Include figure files
\usepackage{dcolumn}% Align table columns on decimal point
\usepackage{bm}% bold math
\usepackage[utf8]{inputenc}
\usepackage[T1]{fontenc}
\usepackage{mathptmx}
\usepackage{etoolbox}
\usepackage{upgreek}
\usepackage{hyperref}
\usepackage{xcolor}
\usepackage[version=4]{mhchem}
\makeatletter
\def\@email#1#2{%
 \endgroup
 \patchcmd{\titleblock@produce}
  {\frontmatter@RRAPformat}
  {\frontmatter@RRAPformat{\produce@RRAP{*#1\href{mailto:#2}{#2}}}\frontmatter@RRAPformat}
  {}{}
}%
\makeatother
\begin{document}

\preprint{AIP/123-QED}

\title{Thermoelectric properties of Bi$_2$O$_2$Se single crystals}
% Force line breaks with \\

\author{Jialu Wang}
      \affiliation{Department of Physics, Zhejiang University, Hangzhou 310027, China}
      \affiliation{Key Laboratory for Quantum Materials of Zhejiang Province, School of Science, Westlake University, Hangzhou 310024, China}
      \affiliation{Institute of Natural Sciences, Westlake Institute for Advanced Study, Hangzhou 310024, China}

\author{Wanghua Hu}
      \affiliation{Key Laboratory for Quantum Materials of Zhejiang Province, School of Science, Westlake University, Hangzhou 310024, China}
      \affiliation{Institute of Natural Sciences, Westlake Institute for Advanced Study, Hangzhou 310024, China}
      \affiliation{Department of Physics, Fudan University, Shanghai 200433, China}

\author{Zhefeng Lou}
      \affiliation{Department of Physics, Zhejiang University, Hangzhou 310027, China}
      \affiliation{Key Laboratory for Quantum Materials of Zhejiang Province, School of Science, Westlake University, Hangzhou 310024, China}
      \affiliation{Institute of Natural Sciences, Westlake Institute for Advanced Study, Hangzhou 310024, China}

\author{Zhuokai Xu}
      \affiliation{Department of Physics, Zhejiang University, Hangzhou 310027, China}
      \affiliation{Key Laboratory for Quantum Materials of Zhejiang Province, School of Science, Westlake University, Hangzhou 310024, China}
      \affiliation{Institute of Natural Sciences, Westlake Institute for Advanced Study, Hangzhou 310024, China}

\author{Xiaohui Yang}
      \affiliation{Key Laboratory for Quantum Materials of Zhejiang Province, School of Science, Westlake University, Hangzhou 310024, China}
      \affiliation{Institute of Natural Sciences, Westlake Institute for Advanced Study, Hangzhou 310024, China}

\author{Tao Wang}
      \affiliation{Key Laboratory for Quantum Materials of Zhejiang Province, School of Science, Westlake University, Hangzhou 310024, China}
      \affiliation{Institute of Natural Sciences, Westlake Institute for Advanced Study, Hangzhou 310024, China}
      \affiliation{Department of Physics, Fudan University, Shanghai 200433, China}

\author{Xiao Lin}
      \altaffiliation {Electronic mail: linxiao@westlake.edu.cn}
      \affiliation{Key Laboratory for Quantum Materials of Zhejiang Province, School of Science, Westlake University, Hangzhou 310024, China}
      \affiliation{Institute of Natural Sciences, Westlake Institute for Advanced Study, Hangzhou 310024, China}

\date{\today}% It is always \today, today,

\begin{abstract}
  Bismuth oxyselenide (Bi$_2$O$_2$Se) attracts great interest as a potential n-type complement to p-type thermoelectric oxides in practical applications. Previous investigations were generally focused on polycrystals. Here, we performed a study on the thermoelectric properties of Bi$_2$O$_2$Se single crystals. Our samples exhibit electron mobility as high as 250~cm$^2.$V$^{-1}$.s$^{-1}$ and thermal conductivity as low as $2$ W.m$^{-1}$.K$^{-1}$ near room temperature. The maximized figure of merit is yielded to be 0.188 at 390 K, higher than that of polycrystals. Consequently, a rough estimation of the phonon mean free path ($\ell_\textrm{ph}$) from the kinetic model amounts to 12 $\textrm{\AA}$ at 390 K and follows a $T^{-1}$ behavior. An extrapolation of $\ell_\textrm{ph}$ to higher temperatures indicates that this system approaches the Ioffe-Regel limit at about 1100 K. In light of the phonon dispersions, we argue that the ultralow $\ell_\textrm{ph}$ is attributed to intense anharmonic phonon-phonon scattering, including Umklapp process and acoustic to optical phonon scattering. Our results suggest that single crystals provide a further improvement of thermoelectric performance of Bi$_2$O$_2$Se.
\end{abstract}

\maketitle

Oxygen-containing thermoelectric (TE) materials have advantages in practical applications owing to their excellent chemical and thermal stability. For instance, p-type BiCuSeO exhibits superior TE performance with a figure of merit ($ZT$) exceeding unity at high temperatures~\cite{Liu2016BiCuSeOZT1.5,Pan2018BiCuSeOZT1.5,Li2012BiCuSeOBadopedZT1.1,Ren2017BiCuSeOZT1.2}. In contrast, n-type oxides show much lower conversion efficiency. In the past decade, bismuth oxyselenide (Bi$_2$O$_2$Se-BOS) was extensively studied as a potential n-type complement to p-type BiCuSeO in TE devices.

BOS is a layered semiconductor, hosting moderate band gap ($\triangle\approx$ 0.8~eV)~\cite{Chen2018SciAdvARPESS}, high electron mobility~\cite{Peng2017ultrahighmobilityNanoLett,WuJinxiong2017NatNano,WangJialu2020NC} and large relative permittivity ($\epsilon_r\approx 155$)~\cite{zhuokai2021permittivity} . Most recently, it was considered to be a promising candidate as a next-generation low-power, high-performance semiconductor. The TE properties of n-type BOS were generally studied in polycrystals with a variety of ion doping, such as Sn~\cite{Nan2015Sn}, Ge~\cite{Nan2018Ge}, Cl~\cite{Nan2017Cl}, Te~\cite{Nan2018Te}, La~\cite{Nan2018La} and Ta~\cite{Nan2019Ta}. Recently, the $ZT$ value is optimized to be as high as 0.69 at about 770 K in BOS polycrystals made from shear exfoliation methods~\cite{Yang2020Sb,Pan2020RecordZT}, owing to the further reduction of the low thermal conductivity ($\kappa$) in this system.
Moreover, gate-tunable power factor as high as 400 $\upmu$W.m$^{-1}$.K$^{-2}$ was reported in atomically thin BOS films \cite{FangYang2020AMGateTE}.

In materials with intrinsic low thermal conductivity, such as SnSe~\cite{Zhao2014SnSeHighZT}, $\kappa$ is significantly restricted by lattice anharmonicity. Single crystals will exhibit outstanding TE performance, due to the enhancement of electrical conductivity ($\sigma$) but without sacrificing low $\kappa$, compared with polycrystals. In BOS, density functional theory (DFT) calculations predicted intrinsic low phonon thermal conductivity because of the unique phonon dispersions~\cite{Wang2018NJPCalZT,Song2019CalKappa,Zhu2019CalLowKappa}.

In this work, we report a study on the thermoelectric properties of BOS single crystals up to 400 K. Slight amounts of electrons are introduced by doping Br ions. The single crystalline materials are metallic with high electron mobility ($\mu_\textrm{300K} \sim$~250 cm$^2.$V$^{-1}$.s$^{-1}$) and low thermal conductivity ($\kappa_\textrm{300K} \sim$~2 W.m$^{-1}$.K$^{-1}$) near room temperature (room-$T$). The maximized $ZT$ value as high as 0.188 at 390 K, exceeding that of polycrystals ever reported. Combined with the specific heat measurement, we roughly estimate that the phonon mean free path ($\ell_\textrm{ph}$) amounts to 12 $\textrm{\AA}$ at 390 K and represents $T^{-1}$ temperature dependence. Extrapolating to even higher temperatures, $\ell_\textrm{ph}$ approaches the Ioffe-Regel limit at about 1100~K. Based on the phonon dispersions from DFT calculations, we discuss two possible scattering mechanisms contributing to the ultralow $\ell_\textrm{ph}$: Umklapp process (U-process) and acoustic ($a$) to optical ($o$) phonon scattering. The study of single crystals reveals an enhancement of TE performance in BOS.

\begin{table*}[!htb]%figure* 表示两列都占 "[]"中为位置参数，四个参数tbph 依次是置顶、置底、浮动、当前位置，，选用的参数优先顺序为h-t-b-p
\centering
  \caption{\label{Tab1} Transport parameters for five BOS single crystals. $n_\textrm{300K}$, $\sigma_\textrm{300K}$ and $\mu_\textrm{300K}$ are the Hall carrier concentration, the electrical conductivity and Hall mobility measured at 300 K, respectively. $T_\textrm{F-Seebeck}$ and $T_\textrm{F-SdH}$ are the Fermi temperature extracted from the slope of thermopower and quantum oscillation measurements from Ref.~\onlinecite{WangJialu2020NC}, respectively. $PF_\textrm{390K}$ is the power factor at 390 K. $ZT_\textrm{390K}$ is the figure of merit at 390 K.}
\begin{ruledtabular}
\begin{tabular}{lccccccccc}

   Samples &No. & $n_\textrm{300K}$  & $\sigma_\textrm{300K}$  & $\mu_\textrm{300K}$ & $T_\textrm{F-Seebeck}$  & ${T_\textrm{F-SdH}}$ & $PF_\textrm{390K}$  & $ZT_\textrm{390K}$ \\
          &   &  $10^{18}\textrm{cm}^{-3}$ & S.cm${^{-1}}$ & cm$^2.$V$^{-1}$.s$^{-1}$ & K & K & $\upmu$W.m${^{-1}}$.K${^{-2}}$ &  \\
  \hline
  Bi$_2$O$_2$Se & 1 & 1   & 81   &507 &  \footnote{~~Refers to quantities that could not be determined.} & 160 & 525 & 0.101 \\
  \hline
  Bi$_2$O$_2$Se$_{1-x}$Br$_x$
  &2 & 6.8 & 321  & 295 & 1892  & 573 & 692 & 0.136  \\
  &3 & 13  & 521  & 251 & 2081  & 882 & 923 & 0.188 \\
  &4 & 22  & 894  & 254 & 3378 & 1253 & 633 & 0.138 \\
  &5 & 87  & 3817 & 274 & 6259 & 3133 & 855 & 0.0788 \\

\end{tabular}
\end{ruledtabular}
%\begin{tabular}{cc}
%\multicolumn{2}{l}{-- ~~refers to quantities which could not be determined.}\\
%\end{tabular}}
\end{table*}

\begin{figure}[!thb]
\begin{center}
\includegraphics[width=9cm]{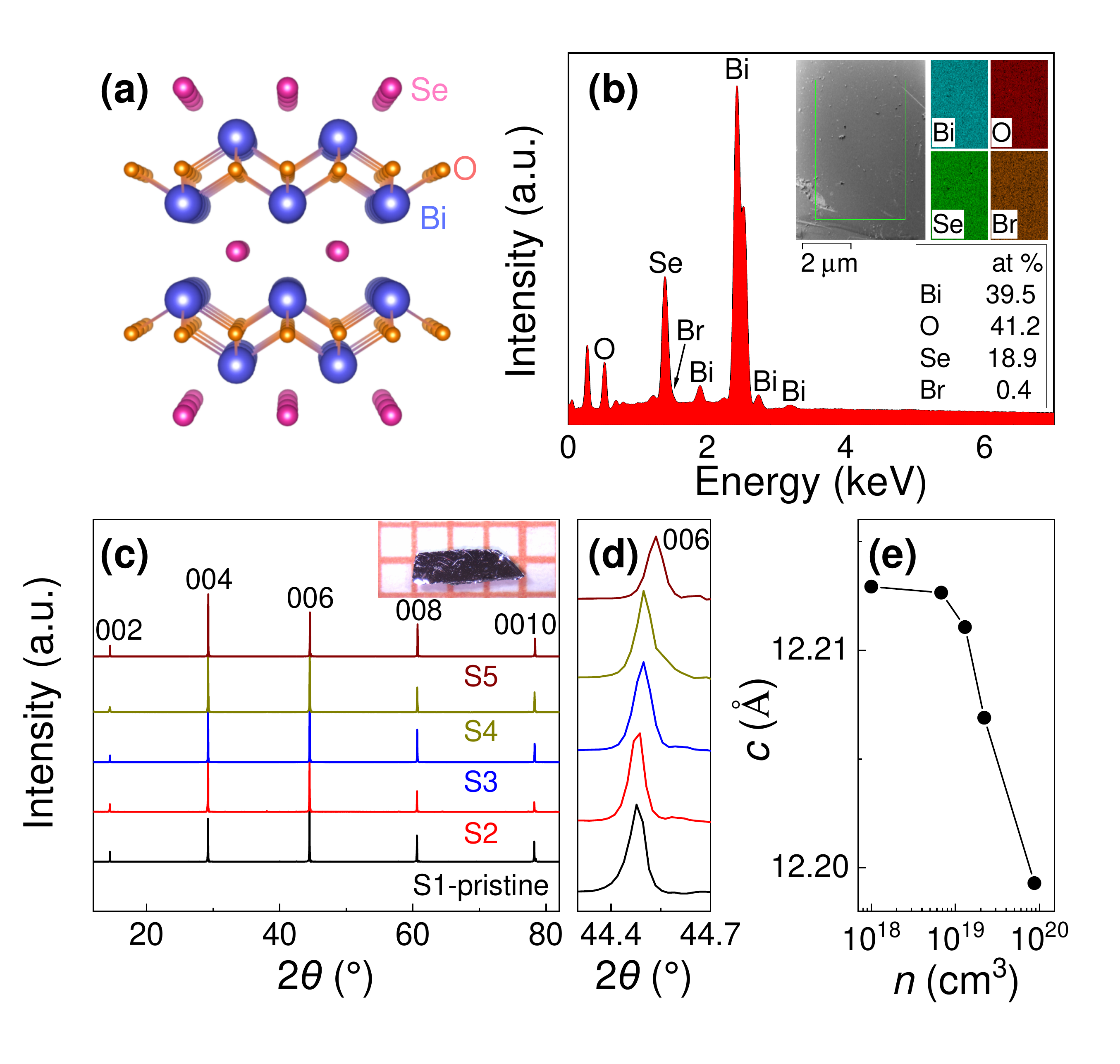}
\end{center}
\vspace{-2.5em}\caption{\label{Fig1}Characterization of BOS crystals. (a) Layered crystal structure of BOS, consisting of alternative stacking of Se and Bi$_2$O$_2$  layers. (b)~EDX spectrum of Br-doped BOS. The insets are the images and element mapping of Bi, O, Se and Br, implying each element is uniformly distributed in the crystal. (c) XRD patterns of BOS single crystals with/without Br-doping on the (001) plane. The inset is a photograph of a Br-doped BOS single crystal. (d) Zoom-in figure around the (006) reflection displaying the evolution of the peak with Br-doping. (e) $c$-axis lattice constant as a function of carrier concentration.}
\end{figure}

Bi$_2$O$_2$Se$_{1-x}$Br$_x$ single crystals were grown through the chemical vapor transport (CVT) method by using poly-crystalline materials as precursors. High purity (5N) Bi, Se, Bi$_2$O$_3$ and BiBr$_3$ powders were mixed and pressed into pellets and then heated in an evacuated quartz ampoule at 823~K for one day.  The resulting polycrystals were thoroughly grounded and sealed in an evacuated quartz tube. The tube was loaded in a horizontal furnace with a temperature gradient from 1123~K to 1023 K over a week. The obtained single crystals are millimeter-size and stable in air.
X-ray diffraction (XRD) patterns were performed by using a Bruker D8 advanced X-ray diffractometer with Cu K$\alpha$ radiation. The composition of samples was determined by an energy-dispersive X-ray (EDX) spectrometer affiliated to a Zeiss field emission scanning electron microscope (SEM). The transport properties and specific heat were measured in Quantum Design PPMS-16T. The thermal conductivity and Seebeck coefficient were measured in a steady-state configuration with one heater and two thermocouples (type-E). All the transport properties were measured along the $ab$ plane. Ohmic contacts were achieved by evaporating gold pads. A photo of the setup for thermoelectric measurements is presented in the supplement. For comparison, a laser flash apparatus (LF467, Netzsch, Germany) was used to obtain the thermal conductivity in a poly-crystalline pellet.
The first-principle calculations were carried out using density functional theory as implemented in Quantum Espresso~\cite{Giannozzi2009QuantumEspresso}. The full phonon spectrum was calculated using the supercell approach as implemented in phonopy~\cite{Togo2015phonopy} with the non-analytical term correction at the $\Gamma$ point included.

As seen in Fig.~\ref{Fig1}a, BOS consists of alternative stacking of Se and Bi$_2$O$_2$ layers, crystallizing in tetragonal phase with $I4/mmm$ space group at room-$T$~\cite{Boller1973structure}. In Fig.~\ref{Fig1}b, EDX data presents the chemical composition of Br-doped BOS: Bi:O:Se:Br = 39.5:41.2:18.9:0.4, indicating slight Br doping. XRD patterns are presented in Fig.~\ref{Fig1}c for BOS single crystals with/without Br-doped at room-$T$. Only $(00\ell)$ reflections shows up, demonstrating the (001) cleavage plane of single crystals. A zoom-in figure around the (006) reflection is shown in Fig.~\ref{Fig1}d. We resolve that the peak shifts to higher angles with Br-doping, implying the $c$-axis contraction, which is consistent with the fact that the radius of Br$^{-}$ is smaller than that of Se$^{2-}$. In Fig.~\ref{Fig1}e, the $c$-axis lattice constant extracted from XRD decreases monotonically by increasing carrier concentration.

\begin{figure*}[!thb]%figure* 表示两列都占 "[]"中为位置参数，四个参数tbph 依次是置顶、置底、浮动、当前位置，，选用的参数优先顺序为h-t-b-p
\begin{center}
\includegraphics[width=7in]{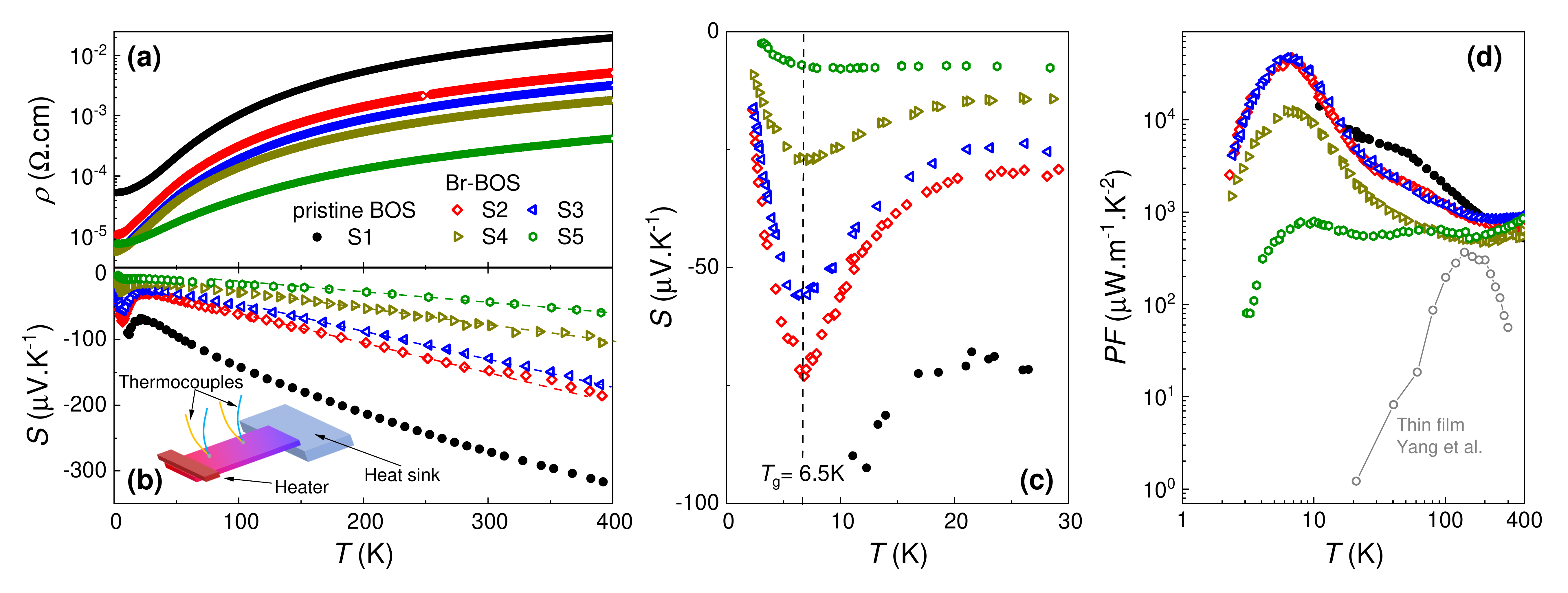}
\end{center}
\vspace{-3em}\caption{\label{Fig2} Electrical and thermoelectric transport properties of BOS single crystals at various $n$. (a) Temperature-dependent resistivity on a semi-log scale for pristine BOS (S1) and Br-doped BOS (S2-S5). (b) Seebeck coefficient as a function of temperature. The dashed lines are fits by Eq.~\ref{S}. Inset is the schematic illustration of the thermoelectric measurement setup. (c) Zoom-in figure of $S$ below 30 K. The dashed lines mark the peak attributed to phonon-drag. (d) Temperature dependence of power factor. The grey open circles are from thin films~\cite{FangYang2020AMGateTE}. [Reproduced with permission from Yang $et~al.$, Adv. Mater. \textbf{33}, 2004786 (2021). Copyright 2021 John Wiley and Sons.]}
\end{figure*}

The temperature-dependent resistivity ($\rho$) for five samples with various Hall carrier concentrations ($n$) is shown in Fig.~\ref{Fig2}a. The transport parameters are summarized in Table~\ref{Tab1}. All the samples are metallic with high electron mobility near room-$T$ ($\mu_\textrm{300K}\approx250-500$~cm$^2.$V$^{-1}$.s$^{-1}$), which is much higher than that of polycrystals ($\mu\approx$ 4.6~cm$^2.$V$^{-1}$.s$^{-1}$ for Cl doped BOS~\cite{Nan2017Cl}, $\mu\approx$ 45~cm$^2.$V$^{-1}$.s$^{-1}$ for Ta doped BOS~\cite{Nan2019Ta} and $\mu\approx$ 190~cm$^2.$V$^{-1}$.s$^{-1}$ for shear exfoliated Sb doped BOS~\cite{Yang2020Sb}). $\rho_\textrm{300K}$ decreases monotonously by increasing $n$ due to Br doping. \ce{Br^-} substitution of \ce{Se^2-} is considered to donate one electron. The defect chemistry reaction can be written as:
\begin{equation}\label{reaction}
\ce{Br^-}\ce{->[\ce{Se^2-}]}\ce{Br_{Se}^$\bullet$}+\textrm{e}^{\prime}
\end{equation}

Figure~\ref{Fig2}b shows the temperature-dependent Seebeck coefficients ($S$) up to 400~K. The negative sign of $S$ demonstrates that electrons dominate the thermoelectric transport in conformity with \ce{Br^-} doping. As seen in the low-temperature zoom-in Fig.~\ref{Fig2}c, $S$ shows a remarkable peak at $T_\textrm{g}\approx6.5$~K for S2, S3 and S4. Note that the peak also exists in S5, which is not as obvious as other samples in this plot because of the small absolute magnitude of $S$ in this specimen. We argue that the peak arises from the phonon-drag effect. In the picture of phonon-drag, $S$ peaks when the characteristic wave vector of phonon ($q_\textrm{ph}$) becomes comparable to the Fermi wave vector ($k_\textrm{F}$), at which phonons are most likely to be felt by electrons. By using the formula $\hbar q_\textrm{ph}v_\textrm{s}=k_\textrm{B} T_\textrm{g}$, where $\hbar$ is the reduced Plank constant, $v_\textrm{s}$ is the sound velocity and $k_\textrm{B}$ is Boltzmann constant, $q_\textrm{ph}$ is deduced to be 0.45~$\textrm{nm}^{-1}$, given $v_\textrm{s}\approx$ 1900~$\textrm{m.s}^{-1}$ \cite{Nan2019Ta}. On the other hand, BOS has an ellipsoid Fermi surface with an anisotropy of 1.8 \cite{WangJialu2020NC}, according to which $k_\textrm{F}$ is estimated to be $0.48~\textrm{nm}^{-1}$, close to $q_\textrm{ph}$, for S2.

At high temperatures, $S$ exhibits a $T$-linear behavior, indicating the diffusion of electrons. The slope ($S/T$) decreases by increasing $n$. In common cases, $T$-linear thermopower occurs at temperature approaching zero K, where temperature independent elastic scattering dominates electron transport. Whereas elastic scattering is only the sufficient condition, but not the necessary. In a few systems, $T$-linear $S$ was observed at high temperatures (e.g. a few times higher than $T_\textrm{g}$) \cite{Ronald1967PRpureAl,Pearson1960goldsilver}, provided that the energy ($\varepsilon$) dependence of electron scattering time ($\tau(\varepsilon)$) does not vary with temperature \cite{Blatt2012Phonondragbook}.

According to the Mott formula, the diffusion term of $S$ is expressed by~\cite{Blatt2012Phonondragbook}:
\begin{equation}\label{}
S=\frac{\pi^{2}}{3} \frac{k_\textrm{B}^{2} T}{\textrm{e}}\left(\frac{\partial \ln \sigma}{\partial \varepsilon}\right)_{\varepsilon_\textrm{F}}
\end{equation}
where $\varepsilon_\textrm{F}$ is the Fermi energy and  $\sigma(\varepsilon)=\frac{\textrm{e}^{2} \ell(\varepsilon) A(\varepsilon)}{12 \pi^{3} \hbar}$ is the electrical conductivity, in which  $\ell(\varepsilon)$ is the energy-dependent mean free path and $A(\varepsilon)\propto\varepsilon$ is the area of constant energy in momentum space. At high temperatures, electron-phonon (e-ph) scattering is overwhelming. Elementary calculations predicted the scattering time $\tau(\varepsilon)\propto\varepsilon^{3/2}$ at $T>T_\textrm{g}$~\cite{Wilsonbook1953}. Note that in elemental metals with large Fermi pockets, $T_\textrm{g}$ approximately approaches the Debye temperature $\Theta_\textrm{D}$. Taking $\ell(\varepsilon)=v(\varepsilon)\tau(\varepsilon)$, in which $v(\varepsilon)\propto\varepsilon^{1/2}$ is the electron velocity, we deduce $\sigma(\varepsilon)\propto\varepsilon^3$. Then the diffusive thermopower at high temperatures is expressed by

\begin{equation}\label{S}
S=\frac{\pi^{2} k_\textrm{B}}{\textrm{e}}\frac{k_\textrm{B}T}{ \varepsilon_\textrm{F}}
\end{equation}
Qualitatively, the slope $S/T$ evolves inversely with $\varepsilon_\textrm{F}$, in consistent with our observations in Fig.~\ref{Fig2}b. Quantitatively,
$T_\textrm{F-Seebeck} = \frac{\varepsilon_\textrm{F}}{k_\textrm{B}}$ extracted from the slope is presented in Table~\ref{Tab1}, which is larger than the value extracted from quantum oscillations by a factor of 2 to 3 \cite{WangJialu2020NC}. A possible reason for the mismatch lies in the fact that BOS has low-lying optical phonon modes that can be easily thermally excited~\cite{WangJialu2020NC,Wang2018NJPCalZT,Song2019CalKappa,Zhu2019CalLowKappa}, which cannot be properly approximated by the simple Debye model (See below).

The TE power factor ($PF=S^2/\rho$) is presented up to 400~K in Fig.~\ref{Fig2}d. For S3, $PF$ amounts to 923~$\upmu$W.m$^{-1}$.K$^{-2}$ at 390 K, increases by lowering temperature and peaks at about 47000~ $\upmu$W.m$^{-1}$.K$^{-2}$. In comparison with polycrystals in previous reports \cite{Nan2015Ag,Nan2015Bideficiencies,Nan2015Sn,Nan2017Cl,Nan2018Ge,Nan2018La,Nan2018Te,Nan2019Ta,Pan2019ShearExfoliation,Ruleova2017Ge,Yang2020Sb,Ruleova2010TE,Pan2020RecordZT}, the power factor of our single crystals is much higher, due to the vast enhancement of electron mobility by suppressing defect scattering prevalent in poly-crystalline materials. The gate-tunable power factors of thin BOS films are presented in Fig.~\ref{Fig2}d \cite{FangYang2020AMGateTE}. Our results from bulk single crystals are apparently better than thin films.

\begin{figure*}[!thb]%figure* 表示两列都占 "[]"中为位置参数，四个参数tbph 依次是置顶、置底、浮动、当前位置，，选用的参数优先顺序为h-t-b-p
\begin{center}
\includegraphics[width=7in]{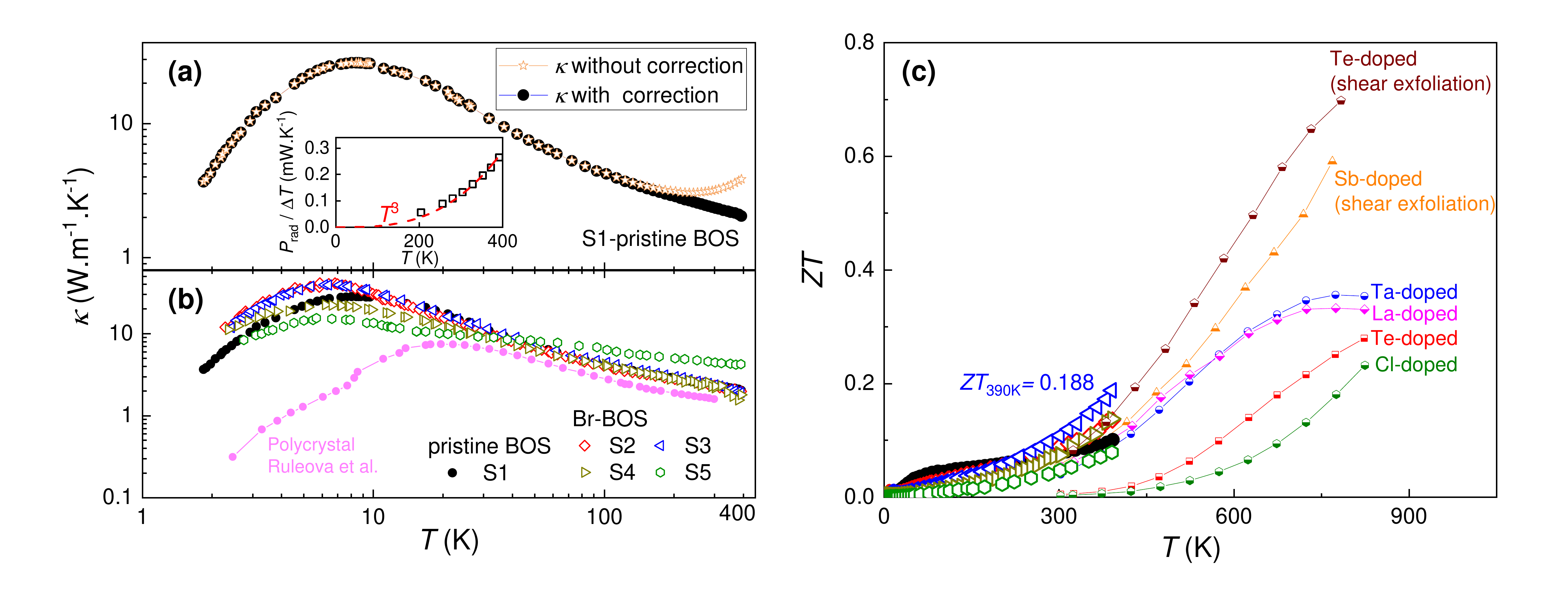}
\end{center}
\vspace{-3em}\caption{\label{Fig3} Thermal conductivity ($\kappa$) and figure of merit ($ZT$) of BOS. (a)  $\kappa$ of pristine BOS (S1) before and after correction. The radiation loss is presented in the inset. The red dashed lines are fits by Eq. \ref{Radiationlossesrevise}. The temperature-independent parameter ($\beta s A$) is 4.52~$\times~10^{-9}$~mW.K$^{-4}$. (b)~$\kappa$~of all five samples after correction. The magenta solid circles are from polycrystals.~\cite{Ruleova2010TE} [Reproduced with permission from Ruleova $et~al.$, Mater. Chem. Phys. \textbf{119}, 299 (2010). Copyright 2010 Elsevier.] (c) $ZT$ as a function of temperature for BOS single crystals, compared with doped polycrystals, including Cl-doped BOS~\cite{Nan2017Cl} [Reproduced with permission from Tan $et~al.$, J. Am. Ceram. Soc. \textbf{100}, 1494 (2017). Copyright 2017 John Wiley and Sons.], Te-doped BOS~\cite{Nan2018Te} [Reproduced with permission from Tan $et~al.$, J. Am. Ceram. Soc. \textbf{101}, 326 (2018). Copyright 2018 John Wiley and Sons.], La-doped BOS~\cite{Nan2018La} [Reproduced with permission from Tan $et~al.$, J. Am. Ceram. Soc. \textbf{101}, 4634 (2018). Copyright 2018 John Wiley and Sons.], Ta-doped BOS~\cite{Nan2019Ta} [Reproduced with permission from Tan $et~al.$, Adv. Energy Mater. \textbf{9}, 1900354 (2019). Copyright 2019 John Wiley and Sons.], shear exfoliated Sb-doped BOS~\cite{Yang2020Sb} [Reproduced with permission from Yang $et~al.$, J. Alloys Compd. \textbf{858}, 157748 (2021). Copyright 2021 Elsevier.], and shear exfoliated Te-doped BOS.~\cite{Pan2020RecordZT} [Reproduced with permission from Pan $et~al.$, Nano Energy \textbf{69}, 104394 (2020). Copyright 2020 Elsevier.]}
\end{figure*}

To quantify the TE performance of BOS, $\kappa$ and $ZT$ are shown in Fig.~\ref{Fig3}. In Fig.~\ref{Fig3}a, the as-measured $\kappa$ of pristine BOS (S1-orange stars) exhibits a broad phonon peak at about 8 K and decreases with increasing $T$ up to 100 K. Further heating leads to an upturn up to 400 K. We may not ascribe this upturn to ambipolar thermal conductivity from thermally excited electron-hole pairs~\cite{Goldsmid1956Bi2Te3Bipolar,Jiang2017Bi2Te3Bipolar}, given the large energy gap ($\Delta\approx 0.8$ eV) of BOS. In the measurements of materials with low $\kappa$, the radiation loss may not be neglected, which will lead to the overestimation of thermal conductivity, especially at high temperatures \cite{2016HaflHeuslar,OuYang2021upturn,Iwasaki2015OrganicSCUpturn}. The radiation loss from the sample to the environment can be estimated by Stefan-Boltzmann law:

\begin{equation}\label{Radiationlossesrevise}
P{\rm{_{rad}}} = \beta s A {T^3\triangle{T}}
\end{equation}
where $\beta$ is the emissivity, $s$ is the Stefan-Boltzmann constant, $A$ is the total radiating surface and $\triangle T$ is the temperature difference between the sample and the surrounding.

By fitting Eq.~\ref{Radiationlossesrevise} to $P{\rm{_{rad}}}$ in the inset of Fig.~\ref{Fig3}a, the temperature-independent parameter $\beta s A$ is extracted (See the supplement for details). As a consequence, the corrected thermal conductivity of S1 is presented in Fig. \ref{Fig3}a (black circles). As seen in the figure, the modification occurs above 200 K. In the supplement, the feasibility of this method is justified by comparing $\kappa$ measured from two different techniques: the steady-state method and the laser flash method, performed in the same polycrystal.

The corrected thermal conductivity for all samples are presented in Fig.~\ref{Fig3}b. Near room-$T$, $\kappa$ of S1-S4 is as low as $2$ W.m$^{-1}$.K$^{-1}$, which only increases by $44\%$ in comparison with polycrystals of previous work \cite{Ruleova2010TE}, despite significantly enhanced crystal quality and reduction of crystal defects in single crystals. Having known $\rho$, $S$ and $\kappa$, the figure of merit $ZT=\frac{S^2 T}{\rho\kappa}$ is shown in Fig.~\ref{Fig3}c, compared with that of doped polycrystals from previous reports \cite{Yang2020Sb,Nan2019Ta,Nan2018La,Nan2018Te,Nan2017Cl,Pan2020RecordZT}. At 390 K, $ZT$ maximizes at $n\approx1.3\times10^{19}$ {cm}$^{-3}$ (S3), which amounts to 0.188, higher than the best performed polycrystal (shear exfoliated Te-BOS: $ZT_\textrm{390K}\approx0.147$) \cite{Pan2020RecordZT}. $ZT_{\rm{390K}}$ of other samples are included in Table \ref{Tab1}. It's reasonable to expect that single crystals will have higher $ZT$ values at higher temperatures compared with polycrystals.

%A rough power-law  extrapolation of $ZT$ of S3 to higher temperatures (the dashed line) reaches a value of 0.8 at about 800 K.

\begin{figure*}[!thb]%figure* 表示两列都占 "[]"中为位置参数，四个参数tbph 依次是置顶、置底、浮动、当前位置，，选用的参数优先顺序为h-t-b-p
\begin{center}
\includegraphics[width=7in]{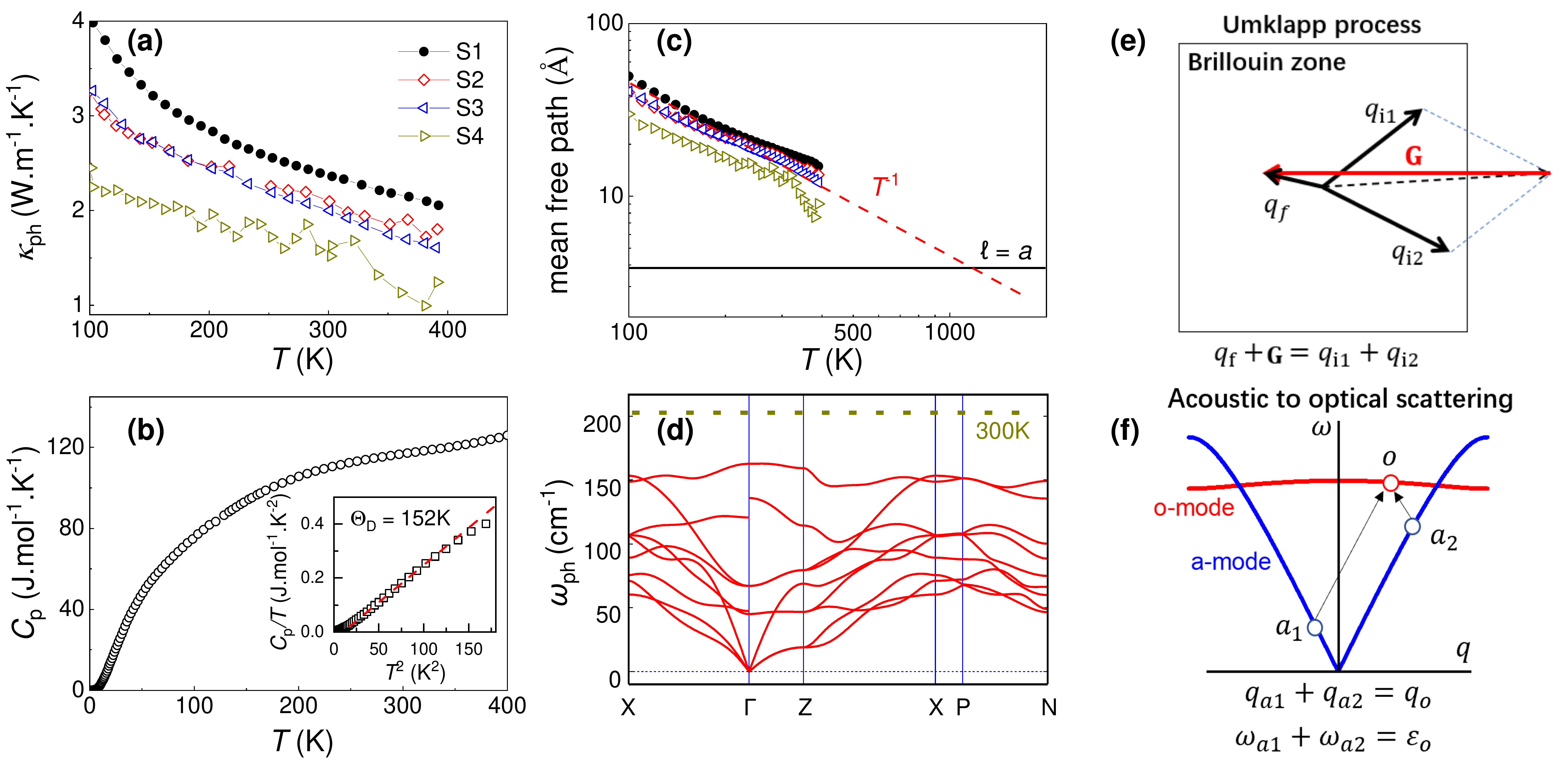}
\end{center}
\vspace{-2em}\caption{\label{Fig4}   Properties of phonons in BOS.  (a) Phonon thermal conductivity ($\kappa_\textrm{ph}$) as a function of temperature above 100 K.  (b) Temperature-dependent specific heat ($C_\textrm{p}$) for pristine BOS (S1). The inset shows $C_\textrm{p}/T$ versus $T^2$ at low temperatures. The dashed line is the fit by the Debye model. (c) Phonon mean free path ($\ell_\textrm{ph}$) as a function of temperature.  The red dashed line marks the behavior of $T^{-1}$. Solid line represents the Ioffe-regel limit ($\ell_\textrm{ph}$ = $a$). (d) Phonon dispersions from DFT calculations. The horizontal dotted line marks the energy of 300~K. (e) Illustration of  U-process of phonons. (f) Illustration of acoustic to optical phonon scattering. }
\end{figure*}

In order to gain insight into the low thermal conductivity of BOS, the phonon contribution ($\kappa_\textrm{ph}$) above 100 K is presented in Fig.~\ref{Fig4}a for S1-S4. $\kappa_\textrm{ph}$ is obtained through the relation: $\kappa_\textrm{ph}=\kappa-\kappa_\textrm{e}$, where $\kappa_\textrm{e}$ is the electronic thermal conductivity estimated from the Wiedemann-Franz law: $L_0=\kappa_\textrm{e}\rho/T$ with $L_0$ the Sommerfeld number (See the details in the supplement).

The temperature-dependent specific heat ($C_\textrm{p}$) for pristine BOS is presented in Fig.~\ref{Fig4}b. The inset shows $C_\textrm{p}/T$ versus $T^2$ at low temperatures, from which the Debye temperature $\Theta_{\rm D}$ is estimated to be 152 {K} through the relation $C_\textrm{p}/T$ = $\gamma$ + $2.4\pi^{4}NR \Theta_{\rm D}^{-3} T^{2}$, where $\gamma$ is the Sommerfeld coefficient, $N$ is the number of atoms per primitive cell and $R$ is the molar gas constant. Moreover, $\gamma$ is negligible, indicating the phonon contribution ($C_\textrm{ph}$) dominates the specific heat in this sample. Given the specific heat of pristine BOS and the average sound velocity $v_\textrm{s}\approx$ 1900 $\textrm{m.s}^{-1}$ \cite{Nan2019Ta}, a rough estimation of the phonon mean free path ($\ell_\textrm{ph}$) can be yielded from  $\kappa_\textrm{ph}$ through the kinetic equation $\kappa_\textrm{ph} = \frac{1}{3}C_\textrm{ph}v\rm{_s}\ell_\textrm{ph}$. As shown in Fig.~\ref{Fig4}c, $\ell_\textrm{ph}$ roughly follows the behavior of $T^{-1}$ for S1-S4 above 100 K. At 390 K, $\ell_\textrm{ph}$ is about 12 $\textrm{{\AA}}$, which is comparable to the out of plane lattice constant ($c$ = 12.2 $\textrm{\AA}$) and threefold the number of in-plane one ($a$ = 3.88 $\textrm{\AA}$). By extrapolation to even higher temperatures, we observed that $\ell_\textrm{ph}$ approaches the Ioffe-regel limit ($\ell_\textrm{ph}$ = $a$) at about 1100 $\textrm{K}$. We note that the kinetic model is oversimplified that may underestimate $\ell_\textrm{ph}$. A more accurate analysis of $\ell_\textrm{ph}$ needs to invoke more sophisticated models concerning phonon dispersions. Overall, these results imply that the ultralow $\ell_\textrm{ph}$ is predominantly determined by anharmonic phonon-phonon (ph-ph) scattering.

Below, let's discuss the fundamentals of the lattice anharmonicity from DFT calculations in this system. The phonon dispersion is presented in Fig.~\ref{Fig4}d, which is similar to our previous reports~\cite{WangJialu2020NC}. First of all, three acoustic modes centered at the $\Gamma$ point disperse at very low energy ( \textless 100 $\textrm{cm}^{-1}$). Secondly, the lowest transverse optical modes (TO) intersect the acoustic modes at about 50 $\textrm{cm}^{-1}$, as shown in SnSe~\cite{Zhao2014SnSeHighZT}, which may lead to strong acoustic-optical interactions. Thirdly, all the phonon modes are thermally excited near room-$T$, providing abundant phase space for ph-ph scattering. The ultralow $\kappa$ is therefore considered to be attributed to the intense ph-ph scattering in BOS.

The well known scattering process in decaying phonon thermal current is the U-process as shown in Fig.~\ref{Fig4}e. We know that $\kappa_\textrm{ph}$ is immune to normal ph-ph scattering, because of the conservation of momentum. According to U-process, the collision of two phonons results in a final state falling out of the first Brillouin zone. In order to go back, the final state loses momentum equal to the reciprocal lattice vector ($\textbf{G}$). The conservation of momentum is allowed to be broken in this way and leads to the reduction of $\kappa_\textrm{ph}$. The U-process requires the wave vector of initial states ($q_\textrm{i1}$ and $q_\textrm{i2}$) larger than a quarter of the smallest reciprocal lattice vector and therefore only manifests itself at high temperatures when large-$q$ phonons are excited.

The process of acoustic ($a$) to optical ($o$) phonon conversion, $a+a\rightarrow o$~\cite{Klemens1966,Yang2017}, also causes a reduction of $\kappa_\textrm{ph}$, as shown in Fig.~\ref{Fig4}f. Though the total energy and momentum are both conserved in this process: $\omega_\textrm{a1}+\omega_\textrm{a2}=\omega_\textrm{o}$ and $q_\textrm{a1}+q_\textrm{a2}=q_\textrm{o}$, where $\omega_\textrm{a1,a2}$  ($q_\textrm{a1,a2}$) are the energy (momentum) of two initial $a$-modes and $\omega_\textrm{o}$ ($q_\textrm{o}$) are the energy (momentum) of the final $o$-mode. The thermal transport efficiency of phonons is suppressed owing to the fact that the group velocity of $o$-modes is much lower than that of $a$-modes. The phase space of $a$ to $o$ modes scattering is limited by the two conservation laws. The larger the hybridization of $a$ and $o$-modes, the larger the phase space, in light of which the $a$ to $o$ scattering rate is expected to be non-negligible in BOS.

In summary, single crystalline BOS shows superior TE properties in comparison with polycrystals. The increase of $ZT$ value is mainly attributed to the enhancement of electrical conductivity as a result of the reduction of scattering by crystal defects. The intrinsic low phonon thermal conductivity is restricted by intense anharmonic ph-ph scattering and shows a lower bound at high temperatures. We argue that in systems with strong lattice anharmonicity, single crystalline materials are generally in advantage of polycrytals in the performance of thermoelectricity.

\section*{Supplementary Material}
See the supplementary material for the photo of thermoelectric setup, the radiation loss correction and the extraction of electron thermal conductivity in Bi$_2$O$_2$Se single crystals.

\begin{acknowledgments}
This research is supported by the National Natural Science Foundation of China via Project 11904294, Zhejiang Provincial Natural Science Foundation of China under Grant No. LQ19A040005 and the foundation of Westlake Multidisciplinary Research Initiative Center (MRIC) (Grant No. MRIC20200402). We thank the support provided by Dr. Chao Zhang and Dr. Ying Nie from Instrumentation and Service Center for Physical Sciences at Westlake University.
\end{acknowledgments}

\section*{Data Availability Statement}

The data that supports the findings of this study are available within the article and its supplementary material.

\section*{References}
%\bibliography{BOS}
%\bibliographystyle{apsrev4-1}


\end{document}